\documentclass[onecolumn,floatfix]{revtex4}
\usepackage{graphicx}
\usepackage{color}
\usepackage{array}
\usepackage{longtable}
\usepackage{amssymb}
\usepackage{amsfonts}
\usepackage{amsmath}

\begin{document}
\title{Do Mathematicians, Economists and Biomedical Scientists Trace Large Topics More Strongly Than Physicists?}
\author{Menghui Li$^{1}$}
\author{Liying Yang$^{2}$}
\author{Huina Zhang$^{1}$}
\author{Zhesi Shen$^{3}$}
\author{Chensheng Wu$^{1}$}
\author{Jinshan Wu$^{3}$\footnote{Corresponding author: jinshanw@bnu.edu.cn}}
\affiliation{
1. Beijing Institute of Science and Technology Information, Beijing 100044, P.R. China \\
2. National Science Library, Chinese Academy of Sciences, Beijing 100190, P.R. China \\
3. School of Systems Science, Beijing Normal University, Beijing 100875, P.R. China
}

\begin{abstract}
In this work, we extend our previous work on largeness tracing among physicists to other fields, namely mathematics, economics and biomedical science. Overall, the results confirm our previous discovery, indicating that scientists in all these fields trace large topics. Surprisingly, however, it seems that researchers in mathematics tend to be more likely to trace large topics than those in the other fields. We also find that on average, papers in top journals are less largeness-driven. We compare researchers from the USA, Germany, Japan and China and find that Chinese researchers exhibit consistently larger exponents, indicating that in all these fields, Chinese researchers trace large topics more strongly than others. Further correlation analyses between the degree of largeness tracing and the numbers of authors, affiliations and references per paper reveal positive correlations -- papers with more authors, affiliations or references are likely to be more largeness-driven, with several interesting and noteworthy exceptions: in economics, papers with more references are not necessary more largeness-driven, and the same is true for papers with more authors in biomedical science. We believe that these empirical discoveries may be valuable to science policy-makers.

\end{abstract}

\maketitle

\section{Introduction}

How researchers choose their research topics has received sustained attention \citep{Busch1983,Diamond1994,Gieryn1978,Merton1938,Zuckerman1978,Foster2015}. This will not only directly affects scientists' output and recognition, but also indirectly affect the science itself. This problem has been widely investigated from sociology of science \citep{Merton1957,Merton1968,Merton1973,Latour1987}, philosophy of science  \citep{Kitcher1995,Strevens2003,Strevens2006,Kleinberg2011,Bikard2015,Boyer-Kassem2015,Zollman2009}, and new economics of science  \citep{Dasgupta1994,Stephan1996}. Quite often previous investigations on this issue are rather qualitative. In this work, we want to study this empirically based on data.

Researchers may determine their research topics primarily according to their research interests, their perceived potential in making progress on the topics or simply by the largeness of the topics, or some combination of all these factors \citep{Foster2015}. Here, largeness refers to how many publications are on the topic during a time interval denoted as $\left[t_{0}, t\right]$. For example, based on knowledge network of chemical reactants, Rzhetsky et al explore the probability of selecting a pair of molecules as a function of the importance (represented by degree) and the difficulty associated with combining them (measured by network distance). It is found that biomedical scientists prefer to explore the local neighborhood of central, important (i.e., higher-degree) molecules in biomedical chemistry  \citep{Rzhetsky2015}. In this work, we focus on the effect of largeness of topics on scientists' choices of topics. It has been found that the scientists tend to trace large topics in physics  \citep{Wei2013}, and in environmental science  \citep{Grandjean2011}.

``The rich get richer"  or Matthew effect is a common social phenomenon  \citep{Barabasi1999,Price1976,Simon1955}. Matthew effects in science have been investigated also in scientometrics, for instance, on scientists' credit \citep{Merton1968} and on citations  \citep{Bonitz1997, Biglu2008, Khosrowjerdi2012}. Only a few previous studies are on scientists' choice of topics in a certain discipline \citep{Wei2013,Rzhetsky2015,Grandjean2011}. Thus, in this paper we ask: Does tracing large topics differ across disciplines? What is the difference of degree of largeness tracing among different countries? To what degree is the intensity of largeness tracing relevant to properties such as the number of authors, references and affiliations of articles? We hope that discoveries from this study will provide valuable information for scientific policy makers especially on the issue of funding and evaluation.

The question of hot topics, prominent topics, or research fronts itself has also been widely investigated in scientometrics  \citep{Boyack2010,Small2014,Upham2010,Chen2004,Chen2006,Cozzens2010,Klavans2016}. Research fronts are generally defined as the areas attracting the most scientific interest in a period of time, especially before publication size of the considered field becomes really big and the field is clearly under exponential growth. Therefore, the studies of hot topics conventionally refer more to identifying emerging hot fields. We will refer these studies as studies on emerging hotness. One might use the number of received citations of papers on a topic \citep{Boyack2010,Small2014,Upham2010,Chen2004,Chen2006,Klavans2016}, or simply use the number of publications (or scientists) on a topic  \citep{Cozzens2010} for this purpose. In this work, we do not need to identify emerging research fronts and our use of hotness or largeness is more like the current size (in terms of number of publications) of the fields. We call the hotness in this sense the realized hotness, or simply largeness.

Furthermore, for each of the four disciplines that we study, which are mathematics, physics, economics and biomedicine, there is an established hierarchical classification system of topics and all the publications under our investigation have been annotated with such a hierarchical codes from the system. Therefore, we do not even need to do classification or clustering of publications to identify topics. In this work, articles with same code at a certain level are considered to be on the same topic. Both the number of papers and the number of received citations of the papers on a topic can be used to quantify the largeness of the topic. However, these two quantities are not really independent: the number of citations more or less follows a power law relation with respect to the number of papers \citep{self-similar,Dong2017}.  In the following analysis, we use only the number of publications belonging to a topic as a measure of the largeness of the topic as for example in \cite{Rzhetsky2015,Grandjean2011}. In principle, one can also consider how new publications are attracted by received citations of topics, but this will be the topic of another investigation. Thus, using established hierarchical classification systems of topics and classifications of papers in these systems, our definition of large topics in this paper is much simpler than that of research fronts: We only need to count how many papers belong to a topic at a given level. The more papers, the larger the topic and the bigger the realized hotness. Using the established hierarchical classification systems of topics is not ideal since new topics emerge constantly. One may use various clustering methods based on citation relation among publications to establish classification systems and identify topics \citep{Waltman:ClusteringComp, Boyack:Clustering, Klavans2016, Waltman:Clustering}. However, in this work, we will use the established coding systems based on controlled vocabularies for the above four disciplines.

Using the above measure of largeness, in this work, we investigate the degree of largeness tracing in mathematics, economics and biomedical science, and then perform a comparison among them and also between these three fields and physics. Intuitively, it might be expected that mathematicians would be more likely to choose their topics of investigation according to their scientific interests and the scientific value of the questions, partly due to the fact that mathematicians intend to work individually and partly since often mathematicians claim so. One might also guess that it is possible that biomedical scientists choose research topics mainly according to medical values of problems rather than the largeness of topics. Economists might choose their topics according to their urgency to the current economic situation rather than their largeness. Here, we empirically examine whether or not this is the case.

Aside from pure curiosity, such a comparison among various fields might have value in the study of science policy. For example, we wish to examine the correlation between the impact of papers and their degree of largeness tracing. As researchers, we would prefer to see that less largeness-driven papers have a greater impact. For science policy makers and administrators of universities, funding agencies and institutes, such a correlation, whether positive or negative, could potentially provide guidance regarding their duties, as one of their goals is to seek strategies to improve the scientific impact of the academic units under their administration. We would also like to examine whether larger teams tend to produce more value-driven papers or more largeness-driven papers. An answer to this question might have strategic policy value regarding how large research teams should be supported.

In our previous work, using data obtained from the American Physical Society (APS) concerning publications in APS journals, we investigated and confirmed the occurrence of largeness tracing in physics  \citep{Wei2013}. In this work, we extended this analysis to the fields of mathematics, economics and biomedical science based on publications from the datasets of the American Mathematical Review (AMR), the Journal of Economic Literature (JEL) and PubMed, respectively. Each of these databases uses a field-specific classification scheme -- the Physics and Astronomy Classification Scheme (PACS) for physics, the Mathematics Subject Classification (MSC) for mathematics, the JEL Classification Codes (JEL) for economics and the Medical Subject Headings (MeSH) for biomedical science -- and classifies each paper into one or more categories, which we call fields or subfields when the categories become finer and finer. When more specific information is needed, we use the following notation, $n$-digit topics, such as $1$-digit topics, $2$-digit topics etc. to denote the fields and subfields at the level of $n$th number/letter of the corresponding classification system. In this sense, we call the $0$th level topics discipline, namely here Physics, Mathematics, Economics and Biomedical science. We test how likely a newly published paper is to be in a large field, where the largeness of each field is measured simply by the current number of published papers in that field accumulated starting from some early years about which we still have publication data. Further details on our methods and data can be found in section \ref{section:data}, where we also define a quantity called the relative contribution ratio, $R^{c}\left(k\right)$, to describe the extent to which researchers from a given country contribute to fields of size $k$ compared with the overall contributions of this country to academia. The exponent $\alpha$ and the relative contribution ratio $R^{c}\left(k\right)$ are the statistics that we analyze throughout the remainder of the paper.

\section{Methods and Data}
\label{section:data}

\textbf{Methods.} Given a set of newly published papers in a time window at time $t$ with size $\Delta t$ ($\left[t, t+\Delta t\right]$), we first record in which fields each paper of them belongs to. We also record the current number of published papers in each field till time $t$ (papers published between $t$ and $t+\Delta t$ are not included) starting from some early years $t_{0}$ about which we still have publication data. Then by combining the above two pieces of information, we count how many of the newly published papers appear in fields of size $k$, which we denote by $p_{k}\left(t\right)$. We are interested in knowing how $p\left(k\right)$ depends on $k$. According to various previous studies \citep{Barabasi1999,Price1976,Simon1955,Wei2013,Rzhetsky2015,Grandjean2011}, often there is a power law $p\left(k\right) \sim k^{\tilde{\alpha}}$. In principle, one can extract this component $\tilde{\alpha}$ by directly fitting $p\left(k\right)$ with respect to $k$.

However, there might be more than one topic with size $k$ (we denote number of size-$k$ topics as $n\left(k\right)$) and often $n\left(k\right)$ also follows a power law with respect to $k$, $n\left(k\right)\sim k^{-\gamma}$. Considering that for each single size-$k$ topic $p\left(k\right) \sim k^{\alpha}$ (this exponent $\alpha$ is what we really want) and there are $n\left(k\right)$ such topics, we will have $p\left(k\right) \sim k^{\alpha}n\left(k\right) \sim  k^{\alpha-\gamma}$. Therefore, if we directly fitting $p\left(k\right)$ to $k$, we would have to know the $\gamma$ of $n\left(k\right)$. In order to get rid of this additional parameter, we define
\begin{align}
T\left(k\right) \triangleq \frac{\frac{p_{k}\left(t\right)}{\sum_{l}p_{l}\left(t\right)}}{\frac{n_{k}\left(t\right)}{\sum_{l}n_{l}\left(t\right)}}.
\end{align}
The normalization is to make $T\left(k\right)$ more like a distribution function. It has been found that in physics this curve is close to a power law \citep{Wei2013}. We then fit this curve of $T\left(k\right)$ to a power law
\begin{align}
T\left(k\right) \sim k^{\alpha}
\end{align}
and refer to the exponent $\alpha$ as the degree of largeness tracing. In fact, to smooth the curves, we use the integral of $T\left(k\right) $ over $k$, which we call the accumulated distribution function $\kappa\left(k\right)$, for the fitting.

Consider a special case where the likelihood of each current paper leading to a new publication is the same, then $p\left(k\right) \sim n\left(k\right) k$ and $T_{k} \sim k$. This is a constant `birth rate' model, which is unlikely to be true since some papers might lead to a lot of new papers while many other papers will not inspire any further publication. Thus $\alpha=1$ is an interesting special case. Another special case is $\alpha=0$ which means that large fields and small fields are equally attractive or inspiring for researchers when they decide to work on which topics. What are the empirical values of the exponent $\alpha$ for various disciplines, or for authors from various countries, might reveal valuable and interesting information about the disciplines and the counties. Therefore, in the following, besides comparing empirical values of $\alpha$ against $1$ and $0$, we will also focus on cross-discipline/cross-countries comparisons, and a few other comparative studies to reveal how other corresponding factors influence largeness tracing.

In addition to the degree of largeness tracing $\alpha$, we also define the relative contribution ratio $R^{c}\left(k\right)$ for each individual country $c$. Given a set of papers and the fields to which they belong, we calculate a quantity $m^{c}_{k}$, which is equal to the number of papers contributed by country $c$ in fields of size $k$ considering all publications from $t_0$ to $t$ in our datasets. In counting $m^{c}_{k}$, we use fractional counting towards countries, in the sense that a country that occurs $y$ times among the $x$ affiliations of a paper is counted as a contribution of $\frac{y}{x}$ from that country. Using this $m^{c}_{k}$, we define
\begin{align}
R^{c}\left(k\right) = \frac{\frac{m^{c}_{k}}{\sum_{\mu}m^{\mu}_{k}}}{\frac{\sum_{l} m^{c}_{l}}{\sum_{\nu, l}m^{\nu}_{l}}}.
\end{align}
Intuitively, the numerator represents the fraction of the contributions made by country $c$ out of all papers in fields of size $k$, whereas the denominator represents the fractional contribution of country $c$ to all fields. More detailed information can be found in \citep{Wei2013}.

\textbf{Data.} In our datasets, each paper is recorded as a data entry that includes the title, year of publication, subject classification code(s), author(s), affiliation(s), reference(s) and number of received citations. The classification schemes in these datasets are all hierarchical. In this work, we use the finest level of each classification system. Articles with the same code are classified into the same topic. Moreover, we regard each paper as one unit of contribution to each assigned topic even when it is assigned to more than one topics. It is not exactly accurate but reasonable for now and we should consider how to evaluate the contribution of each paper to their topics more accurately in future studies. It is found that the size of topics follows a skewed distribution \citep{Wei2013}.

The \textbf{physics} dataset is a collection of all papers published in APS journals from $1976$ to $2013$.  Here, we consider only those research papers, e.g., articles, brief reports and rapid communications, that have been annotated with PACS numbers. In physics, we consider 6-digit topics, e.g., 03.65.-w(Quantum mechanics). We consider a total of $389912$ papers, $5823$ PACS numbers, and $1131566$ classification labels.

The \textbf{mathematics} dataset is a collection of all papers collected by Mathematical Reviews from $1969$ to $2015$. Here, we consider only those journal papers recorded by both Mathematical Reviews and Web of Science (WoS). The MSC codes for each paper are taken from Mathematical Reviews, and other information is from WoS. There are significant differences between different versions of the MSC codes. Thus, we study the mathematics dataset separately for three different time periods, from 1991 to 1999, from 2000 to 2009 and from 2010 to 2015. In mathematics, we consider 5-digit topics, e.g., 40C05(Matrix methods). We have respectively $92992$, $289554$ and $261504$ papers, $5143$, $5124$ and $5487$ MSC codes, with $456994$, $755121$, $711349$ classification labels for the period of 1991-1999, 2000-2009 and 2010-2015.

The \textbf{economics} dataset is a collection of all economics papers collected in the American Economic Association JEL database from $1970$ to $2013$. Here, we consider only those papers recorded by both JEL and WoS. The JEL codes are taken from the JEL database, whereas other information is from WoS. In economics, we consider 3-digit topics, e.g., D12 (Consumer Economics: Empirical Analysis). We consider $241751$ papers, $1125$ JEL codes and $582150$ classification labels.

The \textbf{biomedical science} dataset is from PubMed, which covers a wide range of important journals in the biomedical and life sciences dating back to $1950$. We downloaded the $2015$ baseline version provided by PubMed. We consider only those papers which have been assigned with MeSH codes. MeSH codes is a classification tree and different branches might have different depths. Some branches stop at 12-digit codes such as A01.456.505.580 while others have a depth of 9-digit such as A17.360.296 or depth of 24-digit such as D27.505.696.875.360.276.210.277. In this work, we only report results on the 12-digit topics. At last, we have $21850751$ papers, $6855$ MeSH codes, and $123446643$ classification labels. We have not yet integrated this data with WoS; consequently, many entries have no affiliations and/or no references.

\section{Results}

We first consider sets consisting of all papers published in each year in each of the three disciplines, determine the overall degree of largeness tracing in each discipline in each year, and compare the results with the degree of largeness tracing in physics. We then divide the set of papers in each discipline according to various characteristics of each paper, including whether it was published in a top journal, its country of origin, the number of authors, the number of affiliations and the number of references, and study the correlations between these features and the degree of largeness tracing.

\subsection{Overall, scientists do trace large topics}
\begin{figure}
\centering
\includegraphics[width=0.8\linewidth]{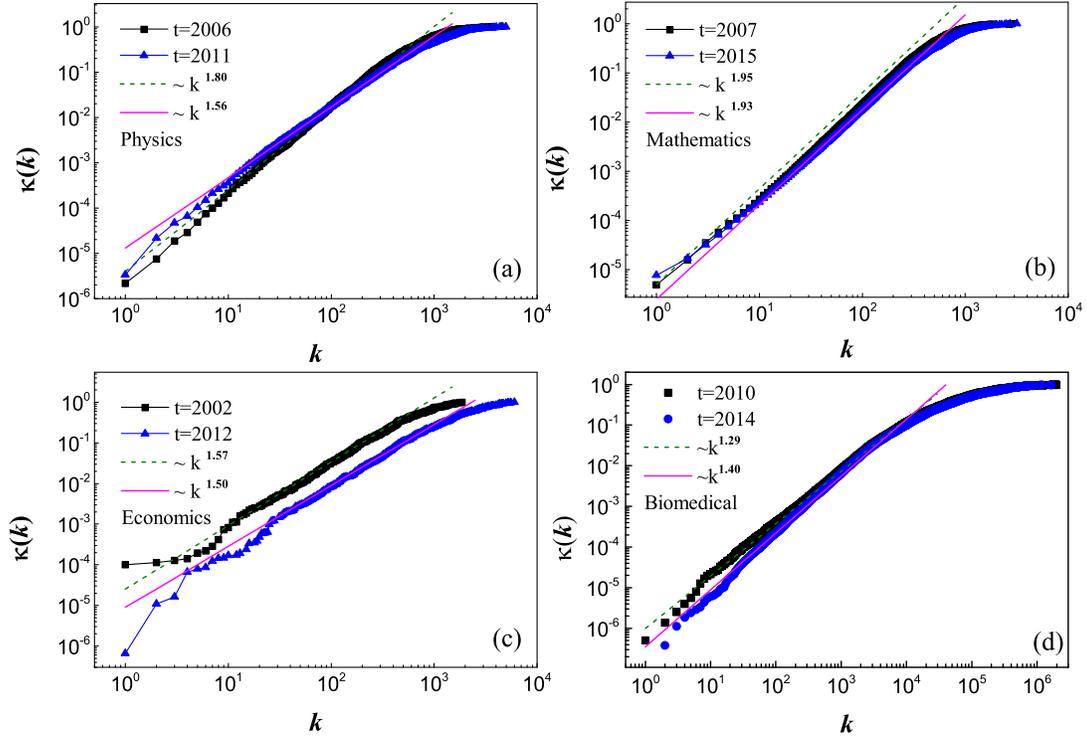}
\caption{\label{overall} (Color online) \textbf{Empirical preferential attachment to large fields for newly published papers}. The cumulative probability functions $\kappa (k)$ for the (a) physics, (b) mathematics, (c) economics and (d) biomedical science datasets show various values of $\alpha$.  (a) is similar to Fig. 1(a) in  \citep{Wei2013} but for different years.}
\end{figure}

\begin{figure}
\centering
\includegraphics[width=0.5\linewidth]{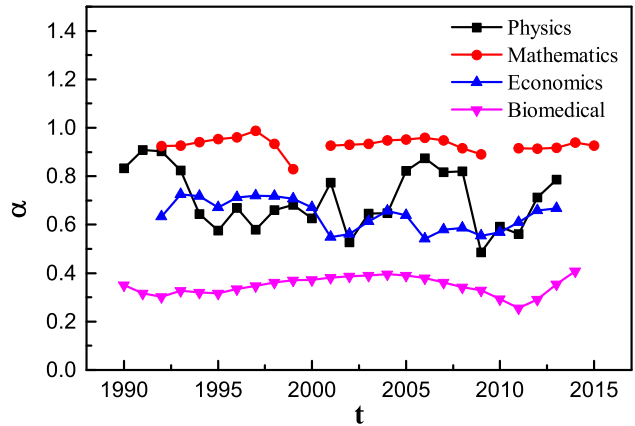}
\caption{\label{exponents} (Color online) \textbf{Evolution of exponents}. The exponents $\alpha$ for different years obtained by least-squares fitting of the curves from $k=10$ to $k=300$ for physics, mathematics and economics and from $k=10$ to $k=3000$ for biomedical science. These fitting ranges were so chosen because the curves deviate from straight lines at large $k$ due to empirically the rarity of such extreme values. The standard deviations of the fitted value of $\alpha$ are smaller than $0.01$. The goodness of regressions $R^2$ are larger than $0.99$. Unless otherwise stated, $\Delta t$ is equal to one year for physics, mathematics and economics and one month for biomedical science. The results for biomedical science are the average values for the corresponding 12 months.  In the year 2000 and 2010, there is a significant change to MSC so that the exponent $\alpha$ cannot be calculated.}
\end{figure}

As shown in Fig. \ref{overall}, the accumulated distribution function $\kappa\left(k\right)$ follows a power law, namely, $\kappa\left(k\right) \sim k^{\alpha +1}$. All of the power-law exponents $\alpha$ are positive, indicating that hotter fields (those that are larger in size) do attract more newly published papers. These results are qualitatively consistent with our observations in physics: generally, scientists are more likely to choose to publish in hotter subfields. The reasons may be as follows. Scientists publish their achievements with high probability by following large topics,  so they may more easily achieve enough recognition to maintain a position. It is also possible that development of science has made a strong correlation between scientific value and largeness of research topics, such that the large topics are the one with most scientific value. In this case, then it is not surprising that overall scientists trace large topics. Unfortunately, it is hard to put this to a test since so far there is no good objective measure of the scientific value of topics. However, differences between the degrees of largeness tracing among the various disciplines are still informative. In particular, the exponent in mathematics is markedly larger than those in the other three fields, as is also shown in Fig. \ref{exponents} for more than $20$ years. This is surprising and counter-intuitive, as it is widely claimed, at least among mathematicians, that the scientific efforts of mathematicians are more independent and more interest-driven or value-driven rather than largeness-driven. The reasons for this discrepancy are not yet known to us. What is also noteworthy is that the exponent in biomedical is less than those in other fields, indicating probably that biomedical scientists are more interest-driven rather than largeness-driven.

\subsection{Influence of journals on largeness tracing: Papers in top journals are less largeness-driven}

\begin{figure*}
\centering
\includegraphics[width=0.5\linewidth]{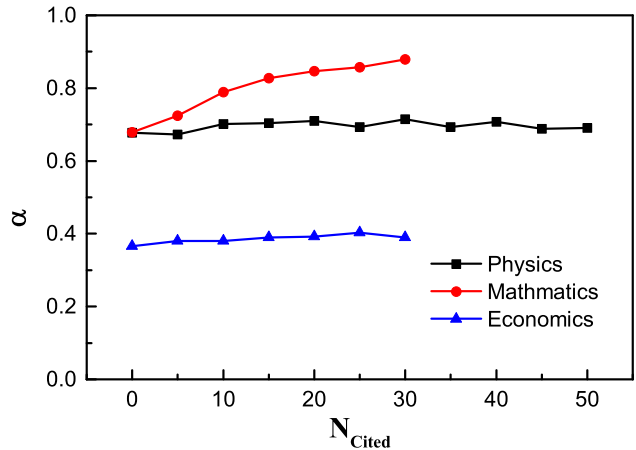}
\caption{\label{citationnumber} (Color online) \textbf{Relation between largeness tracing and cited number}. The preferential attachment exponents $\alpha$ increase slightly with respect to the cited number $N_{Cited}$.}
\end{figure*}

We also like to know whether, on average, largeness-tracing papers have less impact on academia as a whole. To this end, we can investigate the correlation between the exponent $\alpha$ and the number of received citations. As shown in Fig. \ref{citationnumber}, doing so reveals no clear-cut pattern, except that in mathematics, there is an overall trend of an increasing value of the exponent with an increasing number of received citations. However, it is generally accepted that papers that are published in top journals are more likely to be original and have a higher impact. Thus, it is possible that they are less largeness-driven. Here we compare the degree of largeness tracing among papers published in top journals with that among papers published in other journals.

\begin{figure}
\centering
\includegraphics[width=0.8\linewidth]{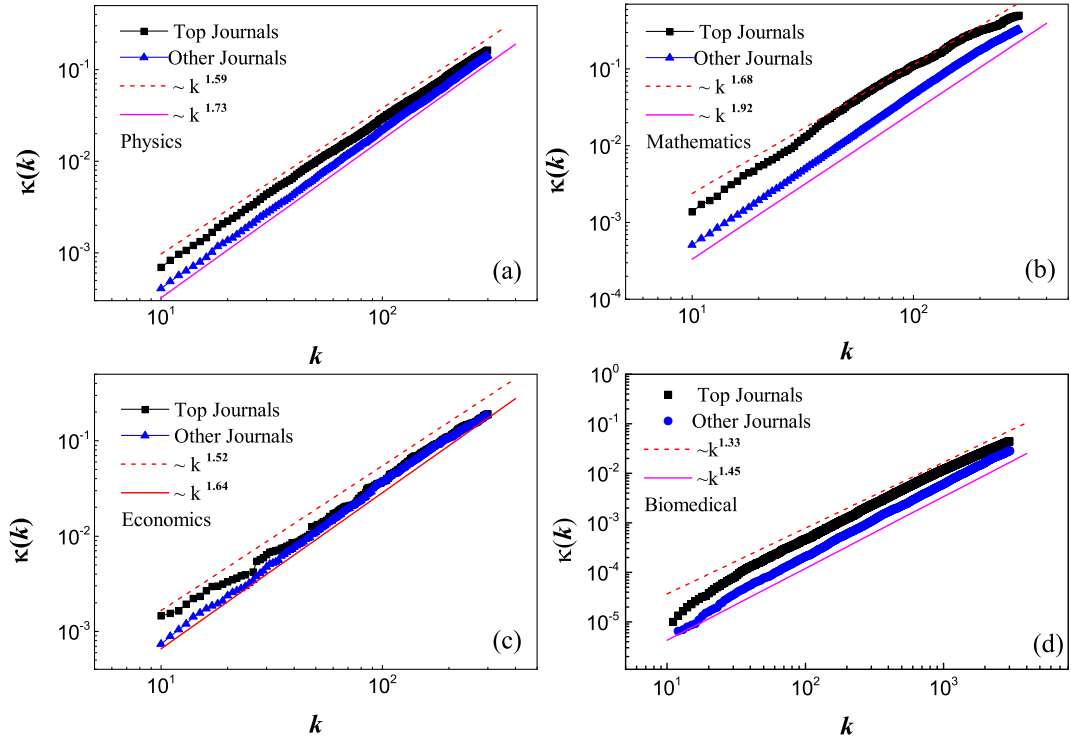}
\caption{\label{citation} (Color online) \textbf{Comparisons of largeness tracing between top journals and other journals}: (a) physics, in which the top journal is Physical Review Letters; (b) mathematics, in which the top journals are Inventiones Mathematicae, Annals of Mathematics, Acta Mathematica, and the Journal of the American Mathematical Society; (c) economics, in which the top journals are the American Economic Review, Econometrica, the Journal of Political Economy, the Quarterly Journal of Economics, and the Review of Economic Studies; and (d) biomedical science, in which the top journals are Nature, Science, the Proceedings of the National Academy of Sciences of the United States of America, Cell, the New England Journal of Medicine and Lancet. }
\end{figure}

We classify the papers into two subsets, i.e., those published in top journals, which are chosen from our databases by disciplinary experts to choose only the highest recognized ones, and those published in other journals, and for each subset, we calculate the exponent $\alpha$. As shown in Fig. \ref{citation}, overall, the exponents $\alpha$ for papers in top journals are smaller than those for papers in other journals in all four disciplines. This finding indicates that publications in top journals are indeed less largeness-driven. The values of the exponent $\alpha$, as indicators of the degree of largeness tracing, are $0.59$ ($0.73$) in physics, $0.68$ ($0.92$) in mathematics, $0.52$ ($0.64$) in economics and $0.33$ ($0.45$) in biomedical science for top (other) journals. We also apply the two-sample Kolmogorov-Smirnov test (KS test) to the sets consisting of the recorded $k$ values of each paper at the time of its publication in top and other journals, and the results of the KS test indicate that for all these disciplines, it is safe to reject the hypothesis that these two sets of $k$ values are drawn from the same underlying distribution. This supports the discovered difference between the exponent $\alpha$ that it is possible that the degrees of largeness tracing of top journal papers are statistically different from that of other papers.

\subsection{Comparison of the degrees of largeness tracing among various countries: Chinese scholars are more largeness-driven}

General public and science policy-makers have a great interest in the research performance of various countries. Here, we compare the degrees of largeness tracing among various countries. We classify the papers into subsets according to the countries that appear in their affiliations. When the authors are from more than one country, the paper is included in the set corresponding to each country. The absolute contribution ratios are listed in table \ref{ACR}. Note that in economics, the contributions from Japan and China are very low. As we will see later, this leads to extremely large fluctuations and, consequently, less reliable observations.

\begin{figure}
\includegraphics[width=\linewidth]{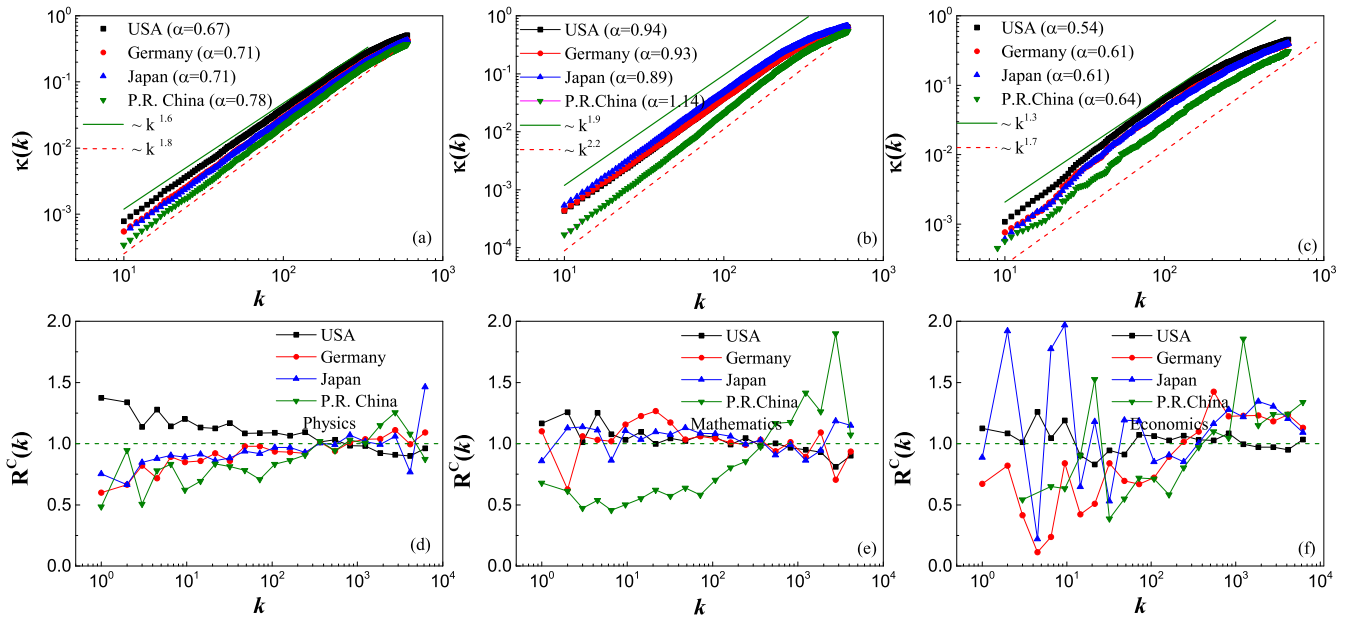}
\caption{\label{countries_overall}(Color online) \textbf{Largeness tracing of scientists in various countries}, as illustrated by the cumulative probability functions $\kappa (k)$ (a) for physics between 1990 and 2012, (b) for mathematics between 2002 and 2008 and (c) for economics between 1995 and 2012 as well as the relative contribution ratios $R^{c}(k)$ for (d) physics, (e) mathematics and (f) economics. The relative contribution ratios $R^{c}(k)$ are grouped together by logarithmic box on the base of $1.5$ and labeled as the intermediate number. (a) and (d) are similar to the corresponding figures in our previous paper  \citep{Wei2013} but with a slightly different calculation of each country's contribution.}
\end{figure}

As seen from Fig. \ref{countries_overall}(a-c), the exponents calculated for China are consistently larger than those for the other three countries in physics, mathematics and economics. Note that the value of Chinese scholars on mathematics is even larger than $1$. Recall that $\alpha=1$ means constant `birth rate' model, where the likelihood of inspiring/attracting/leading to a new publication is the same for each current paper. With this $\alpha>1$, it implies that for Chinese mathematicians it is even possible that larger fields are disproportionately more attractive than the other fields. We do not want to speculate further on why it is the case in China. However, since this observation that Chinese scholars have larger $\alpha$ is there in all the three disciplines (In Economics, it is still larger but not that much larger than the $\alpha$ of others countries), other researchers should look into possible systematic reasons behind it. To further characterize the differences among these countries, we consider the relative contribution ratios $R^{c}\left(k\right)$.

As shown in Fig. \ref{countries_overall}(d-f), in all three of these disciplines, the relative contribution ratio $R(k)$ of China is smaller in cold fields (for small $k$ values) and larger in large fields (for large $k$ values), whereas the opposite is true for the USA. Thus, Chinese scholars make more (fewer) contributions to large (cold) subfields compared with their average contribution to all subfields. These results further indicate that Chinese scientists are more likely to follow large topics. We have not investigated potential institutional reasons for this qualitative difference in behaviour between Chinese and US scholars, although we strongly suspect that causes may exist at a systematic level. In our definition of $R\left(k\right)$, in principle, the numerator $\frac{m^{c}_{k}}{\sum_{\mu} m^{\mu}_{k}}$ should eliminate the influence of the absolute sizes of the fields. Therefore, proportionally speaking, in the absence of any other determining factor, the relative contribution ratio should be approximately constant, as in the case of mathematics in Germany and Japan, as illustrated in Fig. \ref{countries_overall}(e). It should be interesting to investigate the possible reasons behind the observed unevenness of the relative contributions to these fields. An educated guess is that the early development of a research program may be regarded as a catch-up period, during which it is easier to focus more strongly on hotter topics, resulting in smaller contributions in cold subfields and larger contributions in large subfields. If this is the case, then it appears that in the field of economics, research in Germany, Japan and China is much less developed compared with research in the USA. This is partially confirmed by the small percentage contributions from Germany, Japan and China, as seen from Table \ref{ACR}.

\begin{table}
\centering
\caption{Absolute contribution ratios.}
\begin{tabular}{lrccc}\hline
& USA & Germany & Japan & China  \\
\hline
Physics&31.4\%& 9.37\% & 7.93\% &4.26\%\\
Mathematics 2000-2009& 21.17\%& 6.15\%& 7.34\%&9.03\%\\
Mathematics 1991-1999& 25.81\%&6.49\% &7.31\%&3.27\% \\
Economics&	41.03\%& 3.73\%& 1.88\%&1.71\%\\
\hline
\end{tabular}\label{ACR}
\end{table}

\subsection{With respect to the numbers of authors, affiliations and references: The more the stronger, with exceptions}

Next, we investigate the influence of the numbers of authors, affiliations and references on the degree of tracking large topics. We classify the papers into subsets according to the numbers of authors, affiliations and references and calculate the largeness-tracing exponent $\alpha$ for each subset. It has been found that in recent years, the size of teams and the share of multi-university collaborations have been larger than ever, and the work of such teams tends to attract more citations \citep{Adams2012}. In addition, it has also been claimed that research works with many authors or many affiliations are typically more focused on large topics \citep{Wei2013, Adams2012}. We would like to examine the relation between team size and the degree of largeness tracing. Knowing whether work from large teams is more likely to be largeness-driven rather than interest- or value-driven may have policy value with regard to explicit support for the building of large teams.

\begin{figure}
\includegraphics[width=\linewidth]{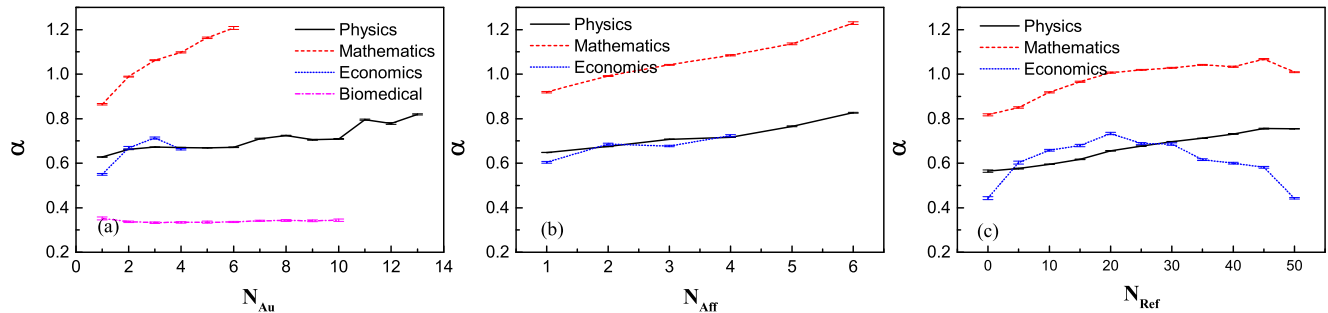}
\caption{\label{author} \textbf{Influence of features on the largeness tracing.} Influences of (a) the number of authors $N_{Au}$, (b) the number of affiliations $N_{Aff}$ and (c) the number of references $N_{Ref}$ on the exponent $\alpha$. The results are shown as the average values of the exponent $\alpha$ for different years, and error bars represent the standard deviations.}
\end{figure}

As shown in Fig. \ref{author}(a), $\alpha$, the exponent quantifying the tracing of large topics, increases with an increasing number of authors, from $0.63$ to $0.82$ in physics, from $0.87$ to $1.2$ in mathematics and from $0.55$ to $0.68$ in economics. These results indicate that large teams typically focus more on large topics, which is consistent with the argument that studies with many authors are typically more focused on large topics  \citep{Adams2012}. Therefore, on average, establishing large teams is not necessarily a good strategy for encouraging value-driven research. However, the exponent $\alpha$ does not increase with an increasing number of authors in biomedical science. This implies that in biomedical science, large teams are not necessary less interest- or value-driven than small teams. In this field, forming large teams might be a good strategy for encouraging value-driven research. Thus, we see that differences between disciplines lead to differences in good team-building strategies. A pure educated guess will be that in the field of biomedical science, there are a significant amount of valuable projects which only large teams can do.

Fig. \ref{author}(b) shows that the exponent $\alpha$ increases with an increasing number of affiliations $N_{Aff}$, from $0.65$ to $0.83$ in physics, from $0.92$ to $1.23$ in mathematics and from $0.6$ to $0.72$ in economics. Thus, the above results indicate that on average, cross-affiliation collaborations are not necessarily a good strategy for encouraging value-driven research.

As seen from Fig. \ref{author}(c), the exponent $\alpha$ also increases with an increasing number of references $N_{Ref}$, from $0.56$ to $0.75$ in physics and from $0.82$ to $1.01$ mathematics, indicating that papers that cite a larger number of references are more likely to address large topics. Interestingly, however, in economics, the exponent $\alpha$ decreases overall with an increasing number of references, from $0.73$ to $0.44$, implying that papers with a larger number of references are more likely to focus on value-driven research topics. The reasons for this exceptional observation regarding the correlation between largeness tracing and the number of references in economics as well as the exceptional correlation between largeness tracing and the number of authors in biomedical science are not yet known to us.

In principle, one can try to adjust $t_{0}$ to see more recent trends as in for example studies of development of social media \citep{socialmedia}. We have performed the same analysis with various $t_{0}$ in measuring sizes of topics. Overall, our above observations on the exponent $\alpha$ and $R\left(k\right)$ still hold for shorter periods between $t_{0}$ and $t$, although their numerical values are different. Moreover, we also have performed the same analysis on various level of MeSH code, such as 6-digit, 9-digit and 15-digit. It is found that the results are qualitatively similar with a different exponent.

\section{Conclusion and Discussion}

In this paper, we investigate the phenomenon of largeness tracing among scientists from four disciplines, namely, physics, mathematics, economics and biomedical science, all of which have yielded a large number of papers with associated subject classification codes. It is found that overall, scientists in all four disciplines are more likely to publish in large fields, but surprisingly, the values of the exponent $\alpha$ are larger for mathematicians than for the others.  We also compare the degrees of largeness tracing between top journals and other journals and find that in all four disciplines, papers published in top journals are less largeness-driven than those in other journals.

In addition, we find that Chinese researchers trace largeness more strongly than US, German and Japanese researchers in all these disciplines. Moreover, it seems that large teams and cross-affiliation collaborations are not necessarily beneficial for encouraging the pursuit of value-driven studies: We find that on average, papers with larger numbers of authors, affiliations and references tend to exhibit larger exponents of largeness tracing, although there are interesting exceptions. Papers with more authors are more largeness-driven in physics, mathematics and economics, whereas in biomedical science, such papers are, in fact, less largeness-driven. It is found that cross-affiliation collaborations are more likely to choose large topics in physics, mathematics and economics. In addition, the degree of largeness tracing increases with an increasing number of references in physics and mathematics, whereas the opposite is true in economics. It is not surprising that papers with large numbers of references are not particularly value-driven or innovative, so this observation in economics is quite non-trivial. The nature of the unique characteristics of biomedical science and economics that cause them to be exceptions to the generally observed trends remains an open and interesting question.

Scientific innovation is an important goal of science. Ideally, we would hope all scientists choose their research topics mainly based on the scientific values of the questions. Of course, it will be even better if scientific value and largeness agree with each other to a large extent. Otherwise, for individual researchers pursuing innovation is risky \citep{Foster2015} and maybe working on large topics is a better strategy. However, at levels of nations or societies, an important goal of science policy is to encourage scientific innovation. For that, we might need a policy to balance innovation and largeness tracing. For example, we have seen that Chinese scholars trace largeness most severely and we speculate that it is related to the fact that researchers got additional bonus payment according to the total number of published SCI papers and also the fact that large teams and big players got a dominantly large portion of national research funds. Publications of large teams, according to our preliminary results, are rather more largeness tracing. In fact, it should be further examined in our future studies whether or not large teams and big players are becoming more and more dominant in China. If it is indeed so, then it in a way helps to make a better sense of our discovered empirical results.

The results presented in this paper are based on the available data. In particular, the investigated physics papers are limited to all papers published by APS journals, which is only a subset of all physics journals. In principle, we should collect all papers published in physics journals, but many journals do not use PACS numbers. Moreover, in PubMed datasets, the affiliation information is often inaccurate and the references are not included, and at present, we have not merged PubMed data with WoS data. In this work, we regard papers published in top journals as high-impact papers. In principle, however, we should consider other measures of impact, such as normalized measures  \citep{Leydesdorff2013, Waltman2013} or measures based on network analysis \citep{Bergstrom2008,Shen2016}. A corresponding study of technological developments could also be interesting, for example, by considering largeness tracing in patents. In this work, we used established hierarchical classification systems to define topics and as we mentioned earlier, clustering methods based on citations might identify different topics and thus lead to different classifications of papers. How various classification methods influence largeness tracing should also be investigated in the future.

\section*{Acknowledgements} This work was partially supported by NSFC Grant 61374175 and Beijing Academy of Science and Technology under project agreement OTP-2014-002. The authors thank APS and PubMed for sharing the data. We highly appreciate both referees' comments and suggestions, which significantly improved the quality of this publication.

\end{document}